\documentclass[pdflatex,sn-mathphys-num]{sn-jnl}

\usepackage{graphicx}%
\usepackage{multirow}%
\usepackage{amsmath,amssymb,amsfonts}%
\usepackage{amsthm}%
\usepackage{mathrsfs}%
\usepackage[title]{appendix}%
\usepackage{xcolor}%
\usepackage{textcomp}%
\usepackage{manyfoot}%
\usepackage{booktabs}%
\usepackage{algorithm}%
\usepackage{algorithmicx}%
\usepackage{algpseudocode}%
\usepackage{listings}%
\usepackage{threeparttable}
\usepackage{tabularx}

\theoremstyle{thmstyleone}%

\theoremstyle{thmstyletwo}%

\theoremstyle{thmstylethree}%

\begin{document}

\title[Development of Automated Data Quality Assessment and Evaluation Indices by Analytical Experience]{Development of Automated Data Quality Assessment and Evaluation Indices by Analytical Experience}
\author*[1]{\fnm{Yuka} \sur{Haruki}}

\author[2]{\fnm{Kei} \sur{Kato}}
\author[2]{\fnm{Yuki} \sur{Enami}}
\author[2]{\fnm{Hiroaki} \sur{Takeuchi}}
\author[2]{\fnm{Daiki} \sur{Kazuno}}
\author[2]{\fnm{Kotaro} \sur{Yamada}}

\author[1]{\fnm{Teruaki} \sur{Hayashi}}


\affil[1]{\orgdiv{School of Engineering}, \orgname{The University of Tokyo}}
\affil[2]{\orgdiv{Kyodo Printing Co., Ltd.}}

\abstract{The societal need to leverage third-party data has driven the data-distribution market and increased the importance of data quality assessment (DQA) in data transactions between organizations. However, DQA requires expert knowledge of raw data and related data attributes, which hinders consensus-building in data purchasing. This study focused on the differences in DQAs between experienced and inexperienced data handlers. We performed two experiments: The first was a questionnaire survey involving 41 participants with varying levels of data-handling experience, who evaluated 12 data samples using 10 predefined indices with and without quality metadata generated by the automated tool. The second was an eye-tracking experiment to reveal the viewing behavior of participants during data evaluation. It was revealed that using quality metadata generated by the automated tool can reduce misrecognition in DQA. While experienced data handlers rated the quality metadata highly, semi-experienced users gave it the lowest ratings. This study contributes to enhancing data understanding within organizations and promoting the distribution of valuable data by proposing an automated tool to support DQAs.}

\keywords{Data Distribution, Data Quality Management, Automated Tool Development}

\maketitle
\section{Introduction}\label{sec1} 
In recent years, data has increasingly become more of a commodity \cite{Schomm13} and the volume of data distributed has grown with the importance of digital transformation \cite{info2020}\cite{stahl16}. Data has also been referred to as ``the oil of the 21st century," and expectations of value creation using data have increased since the global boom in big data around 2013. In line with this, a data-distribution service emerged, where data is traded between companies and organizations \cite{stahl2016classification}.\par
Data-distribution services provide the necessary infrastructure and services to facilitate the exchange of data. This allows information stored in dead storage within individual organizations or companies to be distributed for use by third parties \cite{balazinska11}. Over the past decade, several data-distribution services, such as DAWEX, AWS Data Exchange, JDEX, and Every Sense Pro, have emerged. Data markets may vary in terms of their underlying business model, type of data offered, functionality, market mechanisms, and so on \cite{spiekermann19}\cite{oh2019personal}. The focus on business use of data through inter-industry transactions is growing \cite{DataMarketDesign}. \par
As data distribution becomes more popular, the focus on how to measure the value of third-party data increases \cite{liang18}. The value of data is determined by a variety of factors, and one of the most important factors is data quality. Data quality is the minimum performance and the characteristic that data must fulfill for the use purpose. The importance of data quality has been highlighted for a long time because insufficient data quality reduces the reliability of the analysis results and the quality of data circulating in data markets varies widely \cite{fan2015data}. Therefore, when purchasing and analyzing third-party data in a company, it is necessary to accurately evaluate the quality of the data under consideration before making a purchase decision. However, several issues need to be addressed when evaluating data. \par
First, evaluation indices for data quality have not been standardized, and discussions on data quality indices have been limited to specific fields \cite{kuwahara1994}\cite{manni2021}. However, with the increase in opportunities to handle data from different fields in recent years, universally accepted quality-evaluation indices not limited to specific fields have become necessary. Second, people evaluate each quality-evaluation index differently. DQA is usually performed by viewing raw data and metadata. The evaluators extract the information necessary for quality evaluation and use their own criteria to determine whether the quality is sufficient. Much of this process is left to the standards and judgment of individual evaluators; however, because each evaluator has a different level of experience, skills, and background knowledge, data evaluation varies based on these. Differences in human perception of quality evaluation hinder smooth data transactions. \par
It is crucial to develop tools to solve these data-quality issues \cite{liang24}. This study aims to design a tool for generating quality metadata to facilitate DQA and evaluate their effectiveness. Quality metadata is metadata that is generated based on raw data and is specialized to aid in the comprehension of data quality for data evaluators. The quality of the data is visualized in the form of graphs, tables, or heat maps for each of the evaluation indices. In this study, we established ten practical evaluation indices and performed two experiments to verify the effectiveness of quality metadata. The first was a questionnaire survey of 41 participants with varying levels of experience in handling data on 12 data samples. The second was an eye-tracking experiment observing the behavior of subjects when evaluating data. The results of the experiments revealed that the evaluation of data quality varied depending on the level of experience in data analysis and that quality metadata can help reduce the misinterpretation of raw data. This study contributes to solving the problem of purchase decision-making being inhibited when DQA differs between evaluators in companies and promoting the distribution of valuable data.\par
The remainder of this paper is organized as follows. Section 2 discusses the relevant studies and the novelty of this study. Section 3 describes the quality metadata features and an automatic method of creating metadata. Section 4 describes the DQA experiments using quality metadata. In Section 5, we describe eye-tracking experiments conducted to clarify the cognitive process. Finally, the conclusions and future work are summarized in Section 6.\par

\section{Related Research}\label{sec2}
\subsection{Assessment of data value}
With the emergence of data-distribution markets, there is a need for methods to measure the value of the data. Research on data pricing for personal data is one of the most active research areas.  Several previous studies have proposed a data transaction and pricing scheme for fair pricing of personal data \cite{Li_2013}\cite{Shen16}\cite{oh2019personal}. Although personal-data trading has been criticized for its risks such as information leakage, it plays a major role in activating the data-distribution market and is a highly important topic \cite{personaldata2023}. Some attempts have been made to estimate the value of the training data used for machine learning based on its contribution to learning results using the Shapley value, a game theory concept \cite{jia2019towards}\cite{liu2020dealer}. Mehta et al. proposed a pricing optimization algorithm for dataset trading, in which buyers filter only the data they are interested in \cite{Mehta2021}. Jiao et al. proposed an optimal pricing algorithm that applied a Bayesian profit maximization mechanism to data-analysis services rather than to the data itself \cite{jiao18}. \par However, most of these studies are based on a single metric, monetary value, and the influence of multiple factors that affect the value of data is rarely discussed. In addition, these methods use only mechanical algorithms to measure the value of data and do not involve human judgment or decision-making. However, in actual data transactions, there are many situations in which people view data and make purchase decisions after carefully examining the content and quality. For example, when a company purchases data, a person mainly engaged in data analysis searches for the data to be purchased based on the contents of the data needed and then evaluates the quality of the data by viewing data samples and metadata. \par
When making a purchase decision, it may be necessary to explain the data content and quality to managers who are not engaged in data analysis, and then ask for their approval. In such cases, if the data is subjected to human evaluation only, it may be difficult to reach a consensus because the evaluation varies from person to person. In such situations, automated tools that support DQA considering human cognitive processes rather than evaluating data values using mechanical algorithms are necessary.

\subsection{Data quality assessment indices}
Attempts have been made to identify and standardize universal DQA indices that are not limited to specific fields. In Japan, the Digital Agency published a data quality management guidebook in 2021 \cite{digitalagency2022}, which proposes 15 quality-evaluation indices to efficiently evaluate and manage data: for example, accuracy (written in the correct format and has no errors in the contents), completeness (has all the necessary variables and no missing data), consistency (no contradictions in the contents) and currentness (the information has been updated to the latest version). This guidebook is an important initiative in data quality management as it officially presents universal quality-evaluation indices from the government. On the other hand, the evaluation indices proposed in this guidebook include those that are difficult to judge and those that are not considered very important depending on the purpose or evaluators; therefore, the indices need to be reconsidered before applying them in practice. \par 
Moreover, developing quality-evaluation indices that apply to a wide range of data and fields is crucial for third-party data exchange. These indices must be highly convincing and consider the common understanding of the players involved in data transactions. In this study, we set ten practical quality-evaluation indices for DQA and evaluated their variability based on data-analysis experience.

\subsection{Automated data-quality assessment tools}
Even before the advent of the data-distribution market, data were used in the same field and organization, and the need to manage data quality, an important factor that determines the value of data, was widely recognized \cite{fan2015data}. An attempt was made to automatically evaluate data quality to improve the efficiency of data quality management in specific fields such as meteorology and biology. For example, Anderson et al. created a tool to identify and remove DNA samples and markers that may cause bias in case-control studies in analytical epidemiology. It maintains data quality and increases the reliability of analytical results \cite{anderson2010}. Kuwahara et al. proposed an algorithm to determine the image quality of satellite remote sensing data by setting quality evaluation items such as blurriness, noise, clouds, and shadows \cite{kuwahara1994}. Setiadi compared two measurement tools widely used in image data quality evaluation: peak signal-to-noise ratio and structural similarity index (SSIM) \cite{setiadi21}.\par
However, as data-distribution markets and services have developed, current research faces three main issues. First, these tools have been developed independently in specific fields and only support data in those fields. In the data-distribution market, where a wide variety of data from different fields is traded, more universal automation tools that can handle data from multiple fields are required. Second, the tools proposed by existing research were developed to manage data owned by the organization, most of which have limited purposes for using the data. Therefore, the criteria for DQA in these domains, such as how high the quality needs to be to be sufficient for certain data use, must be strictly standardized. However, in the data-distribution market, it is assumed that data created by other organizations will be used, and the purpose of the data use is not always the same; therefore, it is difficult to determine whether the quality is sufficient for the purpose using only tools that uniformly evaluate the quality. Therefore, information summarization techniques are needed to support human data evaluation, rather than mechanically evaluating different types of third-party data. Third, current DQA tools are intended to be used only by people who are familiar with the data, such as researchers. However, when companies and other organizations trade data, managers who authorize the purchases are not necessarily familiar with the data, and stakeholders have varying experiences in handling data. Thus, support tools that enable people with a wide range of attributes to understand data quality are necessary. \par
In this study, we focused on data quality and proposed an automated tool that provides quality metadata to support DQA. This automated tool is not specialized for data in a specific field, as in the current DQA tools; it enables universal evaluation across fields. Furthermore, by conducting a questionnaire survey using this tool, it became evident how multiple factors influence the personal assessment of data value, as well as the utility and comprehensibility of cognitive process regarding quality metadata, depending on the level of experience of the evaluator when interacting with the data. If data evaluation differs depending on the evaluator, it hinders the decision to purchase data. By using quality metadata, it is possible to reduce the variation in DQA and contribute to promoting data transactions.

\section{Functionality and Implementation of an Automated Data Quality Assessment Tool}\label{sec3}
\subsection{Selection of Quality Assessment Indices}
The quality assessment indices to be implemented in the automation tool were selected, as listed in Table 1, based on discussions with practitioners and reference to previous examples of data quality control \cite{digitalagency2022}. The suitability of the support for automation was also considered. The 10 evaluation indices selected were quantity, accuracy, granularity, completeness, uniqueness, precision, and compliance, which are quality indices related to pure data quality, and rarity, universality, and linkage. These are variable indices considered as the values of variables in the dataset. A variable is an attribute that defines the values contained in a dataset and is equivalent to the column name in the tabular data. In the existing studies \cite{anderson2010}\cite{kuwahara1994}, automated tools have proposed algorithms for evaluation indices that are only necessary for data in specific fields. However, our study comprehensively handles more universal evaluation indices and provides them with quality metadata.

\begin{table}[h]
    \centering
    \caption{Ten evaluation indices for DQA employed in this study}
    \renewcommand{\arraystretch}{1.5}  
    \begin{tabularx}{\linewidth}{@{}lX@{}}\toprule
    Evaluation indices  &  Definition\\ \midrule
       Quantity & Sufficient quantity of data for analysis\\
       Accuracy  & No noticeably erroneous data\\
       Granularity & Granularity of information regarding time series, location, etc. must be Sufficient for analysis\\
       Completeness & Sufficiently small number of missing values (missing data)\\
       Uniqueness & No duplication of data and ability to identify a single piece of data\\
       Precision & Data must be sufficient and complete in terms of significant digits\\
       Compliance & The data format and value units are common and easy to analyze together with other data sets\\
       Rarity & Rarity of the variables included in the data in the relevant data field\\
       Universality & Universality of the variables included in the data in the relevant data field\\
       Linkage & Variables that can be easily linked to datasets from other fields are included\\ \bottomrule
    \end{tabularx}
    \label{tab:my_label}
\end{table}

\subsection{Implementation Details for Data Quality Visualization}
Quality metadata visualizing the quality information of the dataset with the above 10 indices was created, and an automated tool was introduced to support DQA (Fig. 1). For the quality indices, the raw data were read in the CSV format, and the quality information for each index was extracted and visualized. In setting granularity, first, the temporal or geographical interval between each field was calculated for the variables specified by the user as variables of importance. Next, the median of these intervals was calculated and defined as granularity. These granularity values were summarized in a tabular format and output as quality metadata.\par
\begin{figure}[h]
    \centering
    \includegraphics[width=0.8\textwidth]{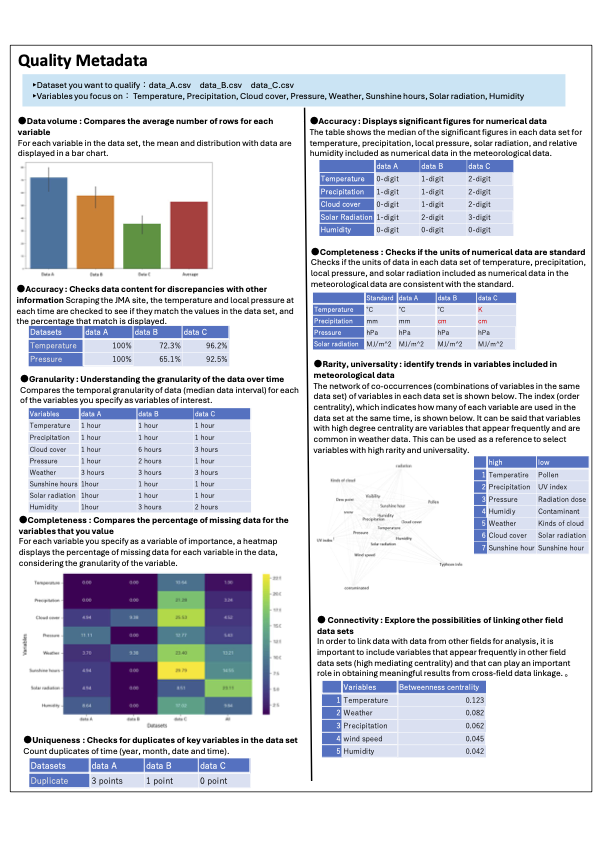}
    \caption{Example of quality metadata}
    \label{fig:enter-label}
\end{figure}

For variable indicators, a variable co-occurrence network including other datasets in the same field as the data to be evaluated was created and visualized. This facilitates an understanding of the relationships between variables in the data. For rarity and universality, variable co-occurrence networks and degree centrality in the networks for datasets in the same field were presented as quality metadata. Jaccard similarity coefficients were used as edge weights to create the networks. Jaccard similarity coefficients are defined by the following equations ($A$ and $B$ are sets of variables):

\begin{equation}
    J(A, B) = \dfrac{|A \cap B|}{|A \cup B|}
\end{equation}

For linkage, we created variable co-occurrence networks using datasets from other fields and presented betweenness centrality as quality metadata. Betweenness centrality is defined by the following equation. Here,
\begin{itemize}
    \item \( B(v) \): Betweenness centrality of vertex \( v \).
    \item \( \sigma_{st} \): Total number of shortest paths from vertex \( s \) to vertex \( t \).
    \item \( \sigma_{st}(v) \): Number of shortest paths from \( s \) to \( t \) that pass through vertex \( v \).
    \item \( V \): Set of vertices in the graph.
\end{itemize}
\begin{equation}
    B(v) = \sum_{s, t \in V} \frac{\sigma_{st}(v)}{\sigma_{st}}
\end{equation}
\par \noindent
Note that the quality metadata is intended for situations in which the quality of multiple datasets is compared considering the purchase of data, and the quality of the multiple datasets to be compared is displayed.

\section{Experiment 1: Data Quality Assessment Experiment}\label{sec4}
\subsection{Purpose} This study aimed to develop an automated DQA tool to support data evaluation and further conduct a questionnaire survey to clarify how the tool influences the decision-making processes of users. Thus, we analyzed the impact of automated tools on DQA using the data-analysis experience categories. The experiment compared the tendency of participants to evaluate the quality of the target data, with and without quality metadata. In addition, because the amount of experience in data analysis may affect DQA, we classified the participants into three categories according to their experience in data analysis: experienced, semi-experienced, and inexperienced. Through these analyses, we confirmed that providing quality metadata when subjects evaluate data quality leads to a correct DQA. In addition, we observed the process by which the participants viewed the raw data and quality metadata to understand the cognitive process of DQA, considering the experience of participants in analyzing data.

\subsection{Datasets}
We used 12 datasets in the experiment: six tourism datasets (A, B, C, G, H, and I) and six meteorological datasets (D, E, F, J, K, and L). We selected these two fields because data in these fields are easily obtained as open data and their characteristics differ with the field. Tourism data are mainly described in natural language and are often used in qualitative analysis. On the other hand, meteorological data contain a lot of numerical data and are often used for quantitative analysis. It is also a data field that requires a lot of domain knowledge to understand the data. These datasets are sample data created by experimenters based on data that have been released as open data, with some additions and modifications. 

In the experiment, we displayed and compared the quality metadata of three datasets from the same field. Therefore, when creating the sample data, the quality of the data was set to change for each of the three datasets to be compared (A, B, C; D, E, F; G, H, I; J, K, L) as shown in Table 2. However, owing to the characteristics of the data, compliance, granularity, and uniqueness of the tourism data were not evaluated.
\begin{table}[h]
    \centering
    \caption{Quality of data used in the experiment}
    \renewcommand{\arraystretch}{1.5}  
    \begin{tabular}{l|ccc|ccc|ccc|ccc} 
        Fields & \multicolumn{3}{c|}{Sightseeing} & \multicolumn{3}{|c|}{Meteorology} & \multicolumn{3}{|c|}{Sightseeing} & \multicolumn{3}{|c}{Meteorology}\\ \hline
        Datasets & A & B & C & D & E & F & G & H & I & J & K & L\\ \hline
        Accuracy & H & L & M & H & L & M & H & L & M & H & L & M\\
        Completeness & M & H & L & M & H & L & L & H & M & M & H & L\\
        Compliance &  &  &  & H & M & L &  &  &  & H & M & L\\
        Precision & L & M & H & L & M & H & M & L & H & L & M & H\\
        Granularity &  &  &  & H & M & L &  &  &  & H & M & L\\
        Quantity & H & M & L & H & M & L & L & H & M & H & M & L\\
        Uniqueness &  &  &  & L & M & H &  &  &  & L & M & H\\
    \end{tabular}
\end{table}

\subsection{Method}
The participants in this experiment were men and women in their 20s to 50s who had experience working in a company, and 90 participants who met these conditions were asked to respond. A total of 41 people responded: 31 employees of the printing company to which some authors belong and 10 students who had participated in long-term internships across various industries, with the types of companies not being specifically designated by us. Participation in the experiment was entirely voluntary, and consent was obtained from the participants regarding the purpose of the experiment and data collection. Additionally, participants were free to withdraw from the experiment at any time at their discretion.

This experiment is primarily an evaluation of the automated tool, and we assume that the influence of the attributes of participants, other than their experience in handling data on the results, is minimal. We specified experience of working in a company as a necessary condition because the automated tool proposed in this study is intended for use in companies. \par
In the experiment, the participants evaluated the data quality according to the ten evaluation indices shown in Table 1. As the evaluation may differ depending on the data-handling experience, the participants were classified beforehand into the following three experience categories. 
\begin{itemize}
    \item Experienced: Those with at least six months of experience in data analysis using programming languages. (12 people)
    \item Semi-experienced: Those who routinely (at least once a month) analyze and view simple data using Excel. (15 people)
    \item Inexperienced: Those who do not routinely have the opportunity to view data. (14 people)
\end{itemize}
\par

For the experiment, we emailed the participants a PDF file containing the raw data and quality metadata to be evaluated and asked them to respond at their convenience. We collected responses by having them fill in and submit a Google Form. For the experiment, we divided the participants into two groups: $\alpha$ and $\beta$ to prevent differences in the data from affecting the evaluation by having the two groups exchange data that were evaluated using only the raw data and data that were evaluated using only the metadata. We ensured that the ratio of experienced and inexperienced participants in each group did not differ significantly based on the preliminary survey on data-handling experience. The time for the experiment was not standardized for each participant, but we asked them to take no more than 2-3 minutes per dataset to answer the questions. The total time required for the experiment was estimated to be 20-30 minutes per participant. No payment was made. The experiment was conducted according to the flow shown in Figure 2.
\begin{figure}[h]
    \centering
    \includegraphics[width=0.7\textwidth]{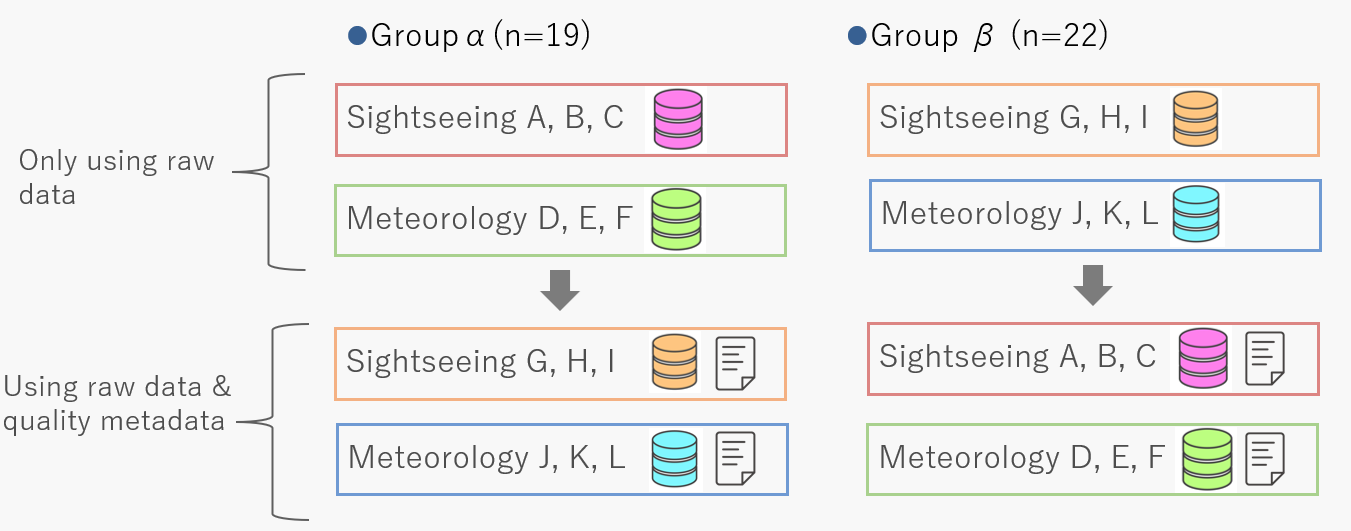}
    \caption{Classification of subject groups and assignment of datasets in the experiment (Group $\alpha$ and $\beta$)}
    \label{fig:enter-label}
\end{figure}

\par
When evaluating the data quality, because quality evaluation may change depending on the purpose of using the data and the data variables required, the purpose of using the data and the variables required to achieve that purpose were presented to the participants as follows.
\begin{itemize}
    \item Sightseeing data\\
    \textbf{Purpose of Data Use}: You are currently involved in the planning and production of a travel magazine. We are considering purchasing tourism data for the publication of a new issue.\\
    \textbf{Variables required}: name, latitude, longitude, address, URL, telephone number, opening hours, days closed, barrier-free information
    \item Meteorological data\\
    \textbf{Purpose of Data Use}: You are currently in charge of purchasing for a retail store. Considering that the demand for products fluctuates according to weather, weather data can be purchased.\\
    \textbf{Variables required}: temperature, precipitation, cloud cover, local air pressure, weather, hours of sunshine, solar radiation, relative humidity
\end{itemize}
The experimental procedure was as follows:
\begin{enumerate}
    \item The 41 participants viewed only the raw data for the three sightseeing datasets and evaluated the data quality for each quality index to determine whether the quality was sufficient for data use. For quality indices, the most appropriate evaluation was selected from the following five levels: ``1. sufficient quality," ``2. slightly insufficient quality, but can be analyzed," ``3. considerably insufficient quality, which may affect the analysis," ``4. insufficient quality, which makes analysis difficult," and ``5. cannot be evaluated." For the ``variable indices," the participants were free to choose variables they considered appropriate (multiple responses were allowed).
    \item As in Step 1, only the raw data of the three meteorological datasets were viewed for quality evaluation. 
    \item The participants viewed the raw data and the quality metadata of the three sightseeing datasets and evaluated them in the same way as in Step 1. 
    \item The participants viewed the raw data and quality metadata of the three meteorological datasets and evaluated the quality of the data as in Step 3.
\end{enumerate}
In addition to this experiment, we conducted an interview survey with two people to find out their intentions and other information that could not be obtained through questionnaires alone.

\subsection{Results and Discussion}
\subsubsection{Evaluability}
First, we analyzed the evaluability of each quality index for each experience category. The ratio of respondents who answered ``cannot evaluate" for the cases where only raw data and quality metadata were viewed is shown in Figure 3.\par
\begin{figure}[h]
    \centering
    \includegraphics[width=\textwidth]{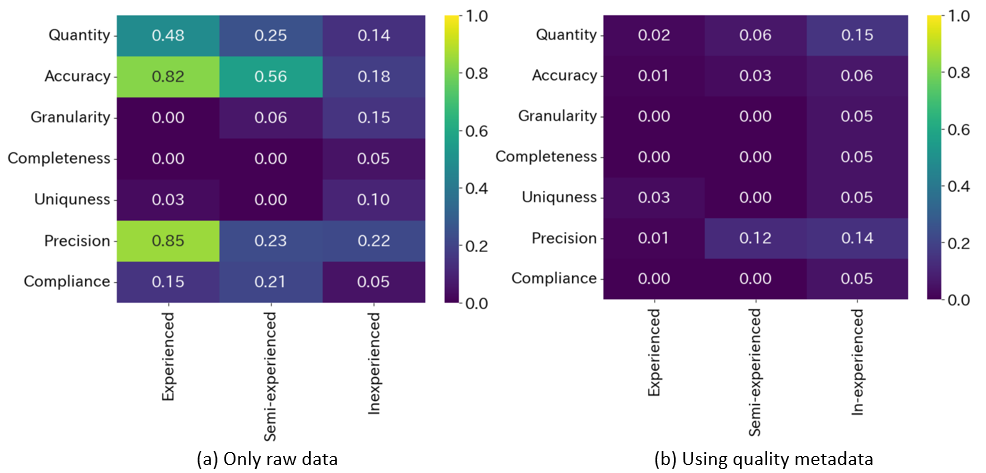}
    \caption{Ratio of subjects who answered ``cannot evaluate."}
    \label{fig:enter-label}
\end{figure}

For each of the three experience categories, a comparison was made between viewing raw data only and viewing the quality metadata. Fisher's exact test confirmed a significant decrease ($p<0.05$) in the number of “cannot be evaluated” answers in terms of quantity, accuracy, compliance (experienced and inexperienced), and accuracy (experienced). For the other indices, the provision of quality metadata also reduces the number of answers. These results suggest that the provision of quality metadata improves DQA. These results clearly show that the provision of quality metadata makes it easier to evaluate the data quality.

\subsubsection{Variation in Evaluation}
Next, we evaluated the variability of the ratings for the quality indices for the four levels of responses other than ``cannot evaluate." First, following \cite{hayashiJSAI2023}, we evaluated the variability in ratings by experience category according to the coefficient of variation (CV). The coefficient of variation is calculated using the following equation, where $\sigma$ is the mean value and $\mu$ is the standard deviation:

\begin{equation}
    CV = \dfrac{\sigma}{\mu}
\end{equation}

\begin{figure}[h]
    \centering
    \includegraphics[width=0.7\textwidth]{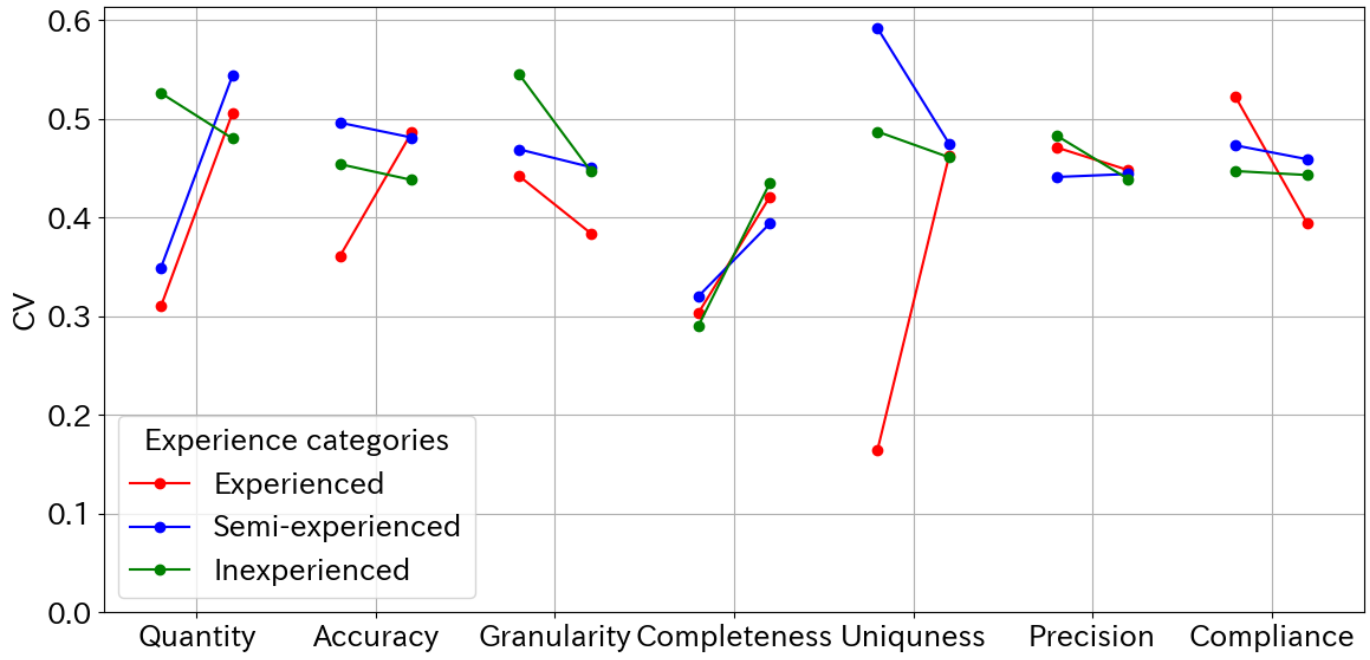}
    \caption{Change in coefficients of variation (CV) with viewing quality metadata}
    \label{fig:enter-label}
\end{figure}

The results are shown in Figure 4. When comparing the results of viewing raw data only and viewing quality metadata for each of the three experience categories, it was found that the provision of quality metadata did not necessarily reduce the variability in judgments about whether the quality was sufficient for the purpose, although the details differed for each evaluation indices and experience category. \par
The results of the error-rate analysis are shown in Figure 5. The false answer rate is the ratio of pairs of data in which the predefined high/low quality of the authors and the high/low quality in the responses are in conflict.\par
\begin{figure}[h]
    \centering
    \includegraphics[width=0.7\textwidth]{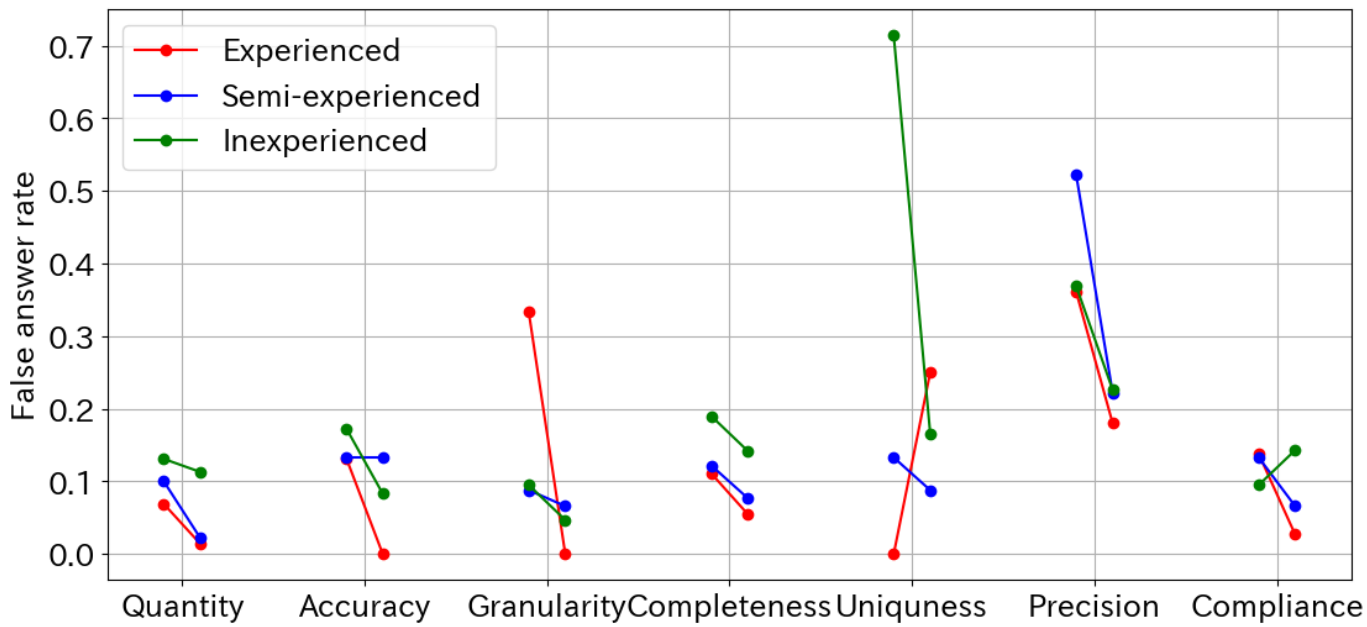}
    \caption{Change in false answer ratio with viewing quality metadata}
    \label{fig:enter-label}
\end{figure}

The error rate was reduced by the provision of quality metadata for almost all indices, except for some indices and experience categories. Fisher's exact test confirmed that the error rate was significantly reduced for all experience categories ($p<0.05$). This is because accuracy is an index that is difficult to evaluate using raw data alone. These results suggest that the provision of quality metadata supports an accurate quality evaluation.

\subsubsection{Variation in Answers to Variable Indices}
Following \cite{saito2015}, we evaluated the variation in the answers to the variable indices using Simpson's diversity coefficient for each data field, and the results are shown in Figure 6. In most indices and experience categories, Simpson's diversity coefficient decreased with the provision of quality metadata. In other words, the provision of quality metadata reduced the variability in the evaluation of the variables in each index. This result suggests that the provision of quality metadata reduces the variability in the evaluation of variable values.

\begin{figure}[h]
    \centering
    \includegraphics[width=0.7\textwidth]{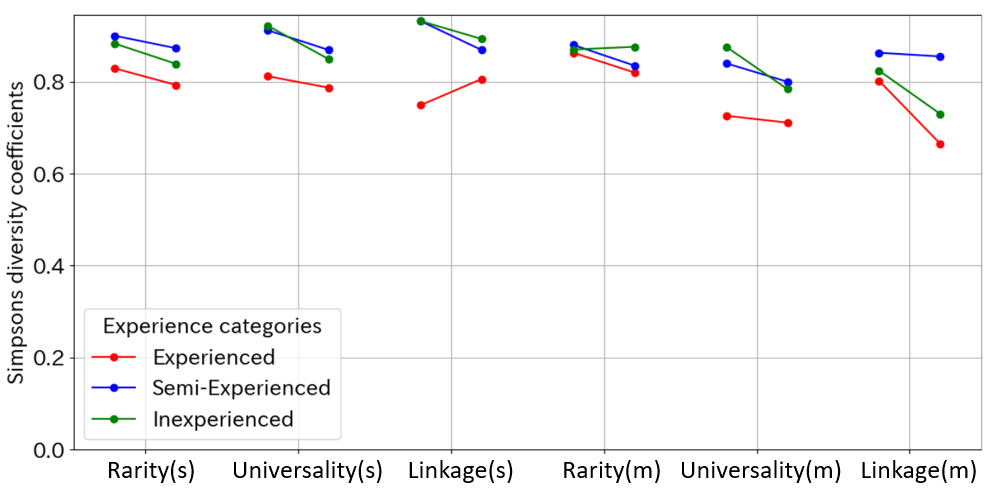}
    \caption{Change in Simpson's diversity coefficients with viewing quality metadata (s: sightseeing data, m: meteorological data)}
    \label{fig:enter-label}
\end{figure}

\subsubsection{Importance of each Index}
A decision tree analysis was conducted based on the evaluation of the quality indices, and the importance of the evaluation indices was clarified. The dataset with the highest quality was classified as ``data to be purchased," and the remaining two datasets were classified as ``data not to be purchased." For each data field, a decision tree analysis was conducted with the score of each evaluation index (however, ``cannot evaluate" was treated as a missing value) as the explanatory variable and ``data to be purchased" or ``data not to be purchased" as the objective variable.\par
Results indicated that completeness is the most important index for the sightseeing and meteorological datasets. Many participants expressed the opinion that completeness was the most important index. The importance of indices other than completeness was compared by data field, and the results were somewhat different. In the interview survey, we obtained opinions that the indices of importance differed depending on the data field.

\subsubsection{Evaluation of the Utility of Quality Metadata}
We asked the subjects to rate the utility of the quality metadata on a five-point scale from ``1. hardly helped" to ``5. greatly helped," and the distribution of ratings by experience category is shown in Figure 7.\par
\begin{figure}[h]
    \centering
    \includegraphics[width=0.7\textwidth]{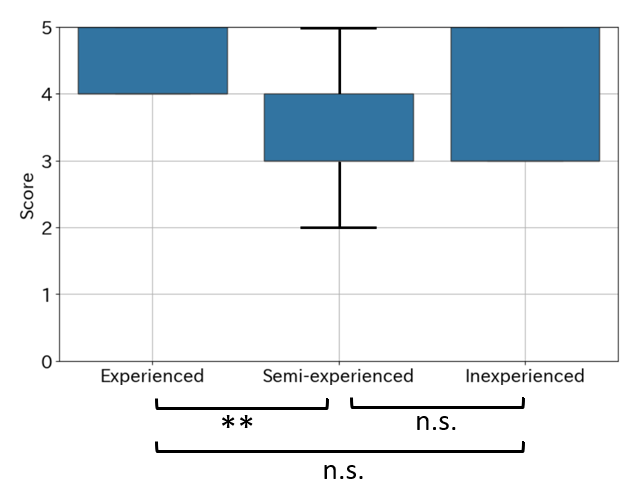}
    \caption{ Evaluation of the utility of quality metadata by participants }
    \label{fig:enter-label}
\end{figure}

First, a Shapiro-Wilk test was conducted to confirm the normality of each of the three experience categories. Results confirmed that the distribution did not follow a normal distribution for the experienced and inexperienced groups ($p<0.05$). The Kruskal-Wallis test, a nonparametric test comparing three or more independent sample groups, revealed that there was a significant difference in the distribution of the three groups ($p = 0.041$). A Steel-Dwass test confirmed that there was a significant difference in the distribution of ratings between the experienced and semi-experienced groups ($p<0.05$).\par
The result suggests that experienced participants tended to rate the quality of metadata highly. On the other hand, the semi-experienced participants tended to rate the quality metadata lowly, and some subjects commented, ``The amount of information increased due to the provision of quality metadata, which was confusing." Inexperienced participants were divided according to their evaluation of metadata. While some highly evaluated the quality metadata as ``easier to evaluate than the raw data," others said, ``There is too much raw data presented in the experiment to begin with, and it stops me from thinking." The analysis results and the interview comments suggest that a certain level of experience in handling data may be necessary to correctly understand quality metadata. In addition, the tendency for semi-experienced participants to lowly rate metadata may be because semi-experienced participants, who often handle data using Excel, tend to place importance on raw data when making quality assessments.

\section{Experiment 2 : Cognitive Process Validation}\label{sec6}
\subsection{Purpose}
In the field of marketing, many studies clarify the cognitive processes of consumers through experiments using eye-tracking devices \cite{KHUSHABA20133803}, but no studies have conducted eye-tracking experiments on data evaluation. By analyzing the frequency of viewing quality metadata and the frequency of viewing each evaluation index in the quality metadata, we obtained insights into the characteristics of viewing behavior and highly important evaluation indices when evaluating data quality using quality metadata.
\subsection{Method}
The participants in this experiment were university students who had long-term internship experience at companies. As in Experiment 1, the participants were divided into three categories according to their experience of handling data. Six subjects in Experiment 2 had previously participated in Experiment 1 and, therefore, had experience in viewing quality metadata. Participation in the experiment was voluntary and unpaid, and the experiment was conducted with the consent of the participants regarding the purpose of the experiment and data acquisition. No personal information was collected. Participants could withdraw from the experiment at their discretion.\par
The participants viewed six meteorological datasets and quality metadata ($3 \times 2$ sets) and selected the dataset with the highest quality for each set. Eye movements were measured while participants viewed the data. We used a Tobii non-contact eye-tracking device (Tobii Pro Spark) to measure eye movement. The Tobii Pro Spark is a 60 Hz eye tracker, suitable for measuring visual attention based on a sampling rate of 60 Hz fixation (place and time of looking). We analyzed the ratio of time spent viewing each index and the number of times the participants looked at the raw data.
\subsection{Results and Discussion}
The results of Experiment 2 are shown in Figure 8. When subjects evaluated the data quality by viewing both the raw data and quality metadata, the semi-experienced group spent more time viewing the raw data (the sum of the stop and saccade times) and moved their eyes most frequently between the raw data and quality metadata. In other words, the tendency to confirm the quality of the raw data was observed even when quality metadata were provided. This characteristic of viewing behavior may have led to the low evaluation of the automatic tool in Experiment 1. The viewing time for each index tended to be longer for completeness, granularity, quantity, and accuracy. All these indices were rated as highly important in Experiment 1. Therefore, it is possible that participants spend more time viewing quality metadata for indices of high importance to improve their understanding of quality.
\begin{figure}[H]
    \centering
    \includegraphics[width=0.9\textwidth]{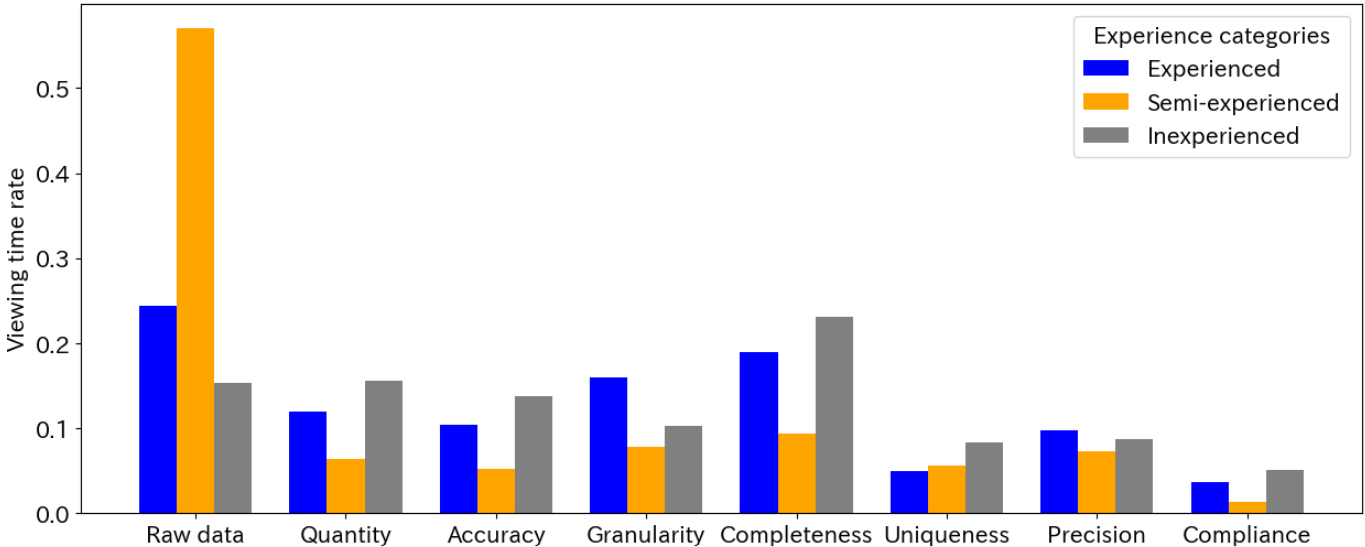}
    \caption{Rate of time spent viewing quality metadata}
    \label{fig:enter-label}
\end{figure}

\par
Based on the above discussion, it can be said that the provision of quality metadata has a certain effect on understanding and evaluating data quality and is particularly effective for experienced users. This is because the experienced users have sufficient ability to understand the content of the quality metadata and can maintain their high level of judgment by appropriately selecting the information they pay attention to when viewing the raw data and quality metadata together. On the other hand, the reason the effect on those with little experience was lower than that on experienced users was that a certain level of experience was required to understand the quality metadata proposed in this study. Many of the semi-experienced and inexperienced users stated that they were tired of looking at the data because they had viewed so much information in the experiment. Thus, it is thought that one of the reasons for the decline in judgment ability was that the lack of experience meant that the information was not sufficiently selected and discarded. Consequently, it is concluded that for semi-experienced and inexperienced users, reducing the amount of information presented when evaluating data quality may be effective in reducing the variation in data quality judgments. For example, it is possible to use only the evaluation indices that were found to be important in this study and provide only quality metadata when evaluating data quality, rather than providing the raw data.

\section{Conclusion }\label{sec5}
In recent years, the emergence of data-distribution markets has provided companies and organizations with more opportunities to buy and sell data. In data transactions, it is essential to accurately evaluate the quality of the data before making a purchase decision. While DQA is typically performed by reviewing raw data or metadata, assessing quality based on raw data is particularly time-consuming and prone to errors. Therefore, support through automated tools is considered highly effective.\par
In this study, we first (1) proposed quality-assessment metrics that can be applied to a wide range of data. We selected ten indices based on existing research and the opinions of practitioners. Next, (2) we created an automated DQA tool, which is unique in that it covers universal DQA indices rather than specialized assessment indices for specific fields, as in existing research. Rather than performing all evaluations automatically, it is necessary for a person to make judgments based on the metadata generated by an automated tool. Finally, (3) we performed an experiment for DQA using an automated tool. \par
The results of the experiment showed that quality metadata supports the accurate evaluation of data quality. By implementing the quality metadata proposed in this study in a data platform, it is possible to facilitate data-distribution by smoothing data quality evaluation and supporting decision-making. In addition, quality metadata was rated highest by experienced participants, while it was rated lowest by semi-experienced participants. This may be because a certain level of experience is required to understand quality metadata, and semi-experienced users tend to value raw data more, as they often manipulate data in Excel. Furthermore, many previous studies use mechanical algorithms to price data and do not fully consider human cognitive processes when evaluating data values. While acknowledging that the data value depends on various factors, it is often evaluated using a single monetary value index. In this study, we evaluated the data using multiple evaluation indices and clarified the importance of each index. Regarding the importance of the evaluation indices, the results showed that completeness is important in all data fields. Further research is required to determine whether this is true for other data fields. \par
Future study prospects include expanding the number of data fields for the experiments. As shown in the experiments, different trends in the DQA and index importance can be observed in other data fields. In addition, it is necessary to reexamine the evaluation indices. \par
In this study, we asked subjects to view quality metadata together with raw data; however, in actual data transactions, it is not always possible to view raw data. Therefore, it is necessary to verify whether it is possible to evaluate the quality by viewing only the quality metadata in the same way as when viewing the raw data. \par
Although the quality metadata proposed in this study supported the understanding of correct data recognition, it was not sufficient to support participants with limited data-handling experience. Therefore, it is necessary to consider ways to support inexperienced users. In addition, the data used in this study is limited to table format. A different approach is required to evaluate the quality of data in different formats.

\bibliography{ref}
\bibliographystyle{junsrt}

\end{document}